\documentclass[fleqn,usenatbib]{mnras}

\usepackage{newtxtext,newtxmath}

\usepackage[T1]{fontenc}

\DeclareRobustCommand{\VAN}[3]{#2}
\let\VANthebibliography\thebibliography
\def\thebibliography{\DeclareRobustCommand{\VAN}[3]{##3}\VANthebibliography}

\usepackage{subcaption}
\usepackage{graphicx}	
\usepackage{amsmath}	
\usepackage{orcidlink}

\title[Silicon-carbides in IRC+10216]{The formation of silicon-carbon chain molecules in IRC+10216 }

\author[T. J. Millar]{
T. J. Millar\orcidlink{0000-0001-5178-3656}\thanks{E-mail: tom.millar@qub.ac.uk}
\\
\\
Astrophysics Research Centre, School of Mathematics and Physics, Queen's University Belfast, University Road, Belfast BT7 1NN, UK\\
}

\date{Accepted 2026 October 23. Received 2026 October 23; in original form 2026 September 29}

\pubyear{2026}

\begin{document}
\label{firstpage}
\pagerange{\pageref{firstpage}--\pageref{lastpage}}
\maketitle

\begin{abstract}
Linear silicon-carbon chain molecules, SiC$_n$ (n = 3--6), have been detected in the circumstellar 
envelope of the carbon-rich AGB star IRC+10216.  Here, we discuss the mechanisms 
likely to be at work in the synthesis of chains up to SiC$_8$ and present for the first time a detailed comparison between  
observations and model calculations.  Using a combination of measured and postulated 
rate coefficients, we show that the column densities of the silicon carbide molecules 
observed in IRC+10216 can be reproduced to with a factor of four.  Furthermore, our model
calculations suggest that the column densities of SiC$_7$ and SiC$_8$ are a factor of 2--4 less 
than that of SiC$_6$. We discuss the influence of neutral-neutral reactions involving parent SiC$_2$
on chain growth and show that their neglect leads to a reduction in column densities of 
carbides up to a maximum of about six for SiC$_6$ and SiC$_7$.  Our results suggest that 
the detection of SiC$_7$ and SiC$_8$ in IRC+10216 will be challenging but that 
astronomical searches should be undertaken.

\end{abstract}




\section{Introduction}
\label{sec:intro}

Silicon-bearing molecules are a common constituent of interstellar clouds and
circumstellar envelopes (CSEs). They range from simple diatomics, such as SiO and SiS, 
detected in many different molecular regions, to more exotic, and perhaps unexpected, species
such as SiCSi \citep{cer15} and CH$_3$SiH$_3$ \citep{cer17}, both 
detected in the inner wind zone of the C-rich AGB star IRC+10216.  More recently, three 
linear silicon carbon chains, SiC$_3$ and SiC$_5$ \citep{cer25} and SiC$_6$ 
\citep{par25} have been detected in IRC+10216. In this 
object, all linear molecules SiC$_n$, n = 3--6, are observed in the outer CSE. 

In recent high-temperature experiments, supported by extensive theoretical studies, 
\citet{taj26} have investigated the formation of silicon carbide grains. 
They also developed a chemical kinetic model that describes the growth of the grain particles from
an initial composition of atomic Si and C, as well as H$_2$ under the experimental conditions 
of high density (n(H$_2$) $\sim$ 10$^{10}$ -- 10$^{12}$ cm$^{-3}$) and temperature (500 K).

In this paper, we investigate a chemical model of the external CSE of IRC+10216 which is 
cooler and less dense region than the dust formation zone. We extend the chemistry of silicon 
beyond that in the RATE22 version\footnote{https://www.umistdatabase.uk} 
of the UMIST Database for Astrochemistry (UDfA) \citep{mil24}, and that used previously 
in our study of the calcium-bearing cyanopolyynes in IRC+10216 \citep{mil26}, to incorporate 
the silicon chain chemistry up to SiC$_8$. In particular, we include the experimental 
results of \citet{par92} on the reactions of SiC$_n^+$ ions with acetylene, C$_2$H$_2$, a
study that has been overlooked in reaction databases to date.

\section{Physical and Chemical Model}
\label{model}

\subsection{Initial conditions}
\label{initial}

Our physical model for the CSE is a simple, one-dimensional, smooth outflow at constant 
velocity and mass-loss rate, using the same values as presented by \citet{mil24}. 
We calculate fractional abundances and radial column densities from an 
initial radius of r$_i$ = 10$^{15}$ cm to an outer radius r$_o$ = 10$^{18}$ cm.  We adopt
the same initial molecular abundances relative to H$_2$ as in \citet{mil24} with  
important differences to the initial abundances of Si-bearing `parent' species, as shown 
in Table \ref{tab:abundances}.  This leads to a reduction of just over an order of magnitude
to the total Si abundance from that in \citet{mil24}.  

\begin{table}
	\centering
	\caption{Initial fractional abundances of Si-bearing species relative to H$_2$. }
	\label{tab:abundances}
	\begin{tabular}{lccl} 
		\hline
		Species & RATE22 & This work & Reference \\
		\hline
		SiO &  5.02 $\times$ 10$^{-6}$ & 2.60 $\times$ 10$^{-7}$ & \citet{mas19} \\
		SiS & 5.98 $\times$ 10$^{-6}$  & 1.30 $\times$ 10$^{-6}$ &  \citet{mas19}\\ 
		SiC$_2$ &  1.87 $\times$ 10$^{-5}$ & 3.70 $\times$ 10$^{-7}$ & \citet{mas18} \\
		SiH$_4$ & 2.20 $\times$ 10$^{-7}$  & 2.20 $\times$ 10$^{-7}$ & \citet{agu20} \\
		\hline
	\end{tabular}
\end{table}

The fractional abundance of SiC$_2$ is set to 3.7 $\times$ 10$^{-7}$
as determined by \citet{mas18} at 2 $\times$ 10$^{14}$ cm, in the region just 
outside the dust condensation zone \citep{agu12}, rather than the much larger value of 
1.87 $\times$ 10$^{-5}$ in \citet{mil24}.

\subsection{Chemical network}
\label{sec:network}

The silicon chain chemistry has been expanded from that in RATE22, as shown in
Table \ref{tab:moldata}, which compares for each species the number of reactions in RATE22 to those in the current network, as well as the values of their electric dipole moments. 
Due to the lack of thermodynamic information and measurements of reaction rate coefficients, products, 
and branching ratios, our chemical network is unable to distinguish between the different isomeric forms of
linear and cyclic silicon carbides, specifically between linear and rhomboidal 
SiC$_3$. To calculate rate 
coefficients for the ion-neutral (IN) reactions of SiC$_3$, we have used the dipole moment 
of the linear form which is about 15 per cent larger than that of the rhomboidal form.

Much of the ion-molecule data describing silicon-carbon chemistry comes from the  
measurements by \citet{wlo91}, who made a comprehensive experimental study of the primary 
and secondary reactions of Si$^+$ with the hydrocarbons C$_2$H$_2$, C$_2$H$_4$, C$_2$H$_6$, 
CH$_3$CCH, CH$_2$CCH$_2$ and C$_4$H$_2$, and from the theoretical chemical kinetic models 
of TMC-1 \citep{her89} and IRC+10216 \citep{mac99}. Importantly for this study, \citet{par92} 
has determined rate coefficients, products and branching ratios for the reactions:

\begin{eqnarray}
  \rm{SiC_n^+ + C_2H_2} & \rightarrow & \rm{products}      
\end{eqnarray}
for n = 2--7.  
Rate coefficients range from (3--6) $\times$ 10$^{-10}$ cm$^3$ s$^{-1}$, with
major product channel being SiC$_{n+2}$H$^+$ + H,  with a minor channel, SiC$_2$H$^+$ + C$_n$H,
occurring for n = 2, 3. The rate coefficient for the Si$^+$ + C$_2$H$_2$ reaction has been measured by 
\citet{wlo91}. In passing, we note that \citet{par92} has also studied the reactions
of Si$_2$C$_n^+$ with C$_2$H$_2$ for n = 1--6.  Although Si$_2$C (SiCSi) has been 
detected in IRC+10216 \citep{cer15}, and is present from the dust formation zone to the outer CSE, 
it is not included in the UDfA network as it is likely formed by LTE chemistry in 
the inner envelope \citep{cer15}.

\begin{table}
	\centering
	\caption{We tabulate the number of reactions included for each species in RATE22 and in the
	current extended version. The dipole moments, with source references, are given in Debye. }
	\label{tab:moldata}
  	\begin{tabular}{lcccl} 
		\hline
		Species & RATE22 & RATE22(ext) & $\mu_D$ & Reference \\
		\hline
		SiC$_2$ & 21 & 35 & 2.39 & \citet{sue89} \\
		SiC$_3$ & 17 & 35 & 4.80 & \citet{mcc00} \\
		SiC$_4$ & 8  & 30 & 6.46 & \citet{gor00} \\
		SiC$_5$ & 0  & 28 & 6.60 & \citet{mcc00} \\
		SiC$_6$ & 0  & 27 & 8.25 & \citet{gor00} \\		
		SiC$_7$ & 0  & 25 & 7.29 & \citet{din00} \\
		SiC$_8$ & 0  & 21 & 10.0 & \citet{bot00} \\
		\hline
	\end{tabular}
	\end{table}

In particular, we include IN reactions of the silicon carbides with H$^+$, 
H$_3^+$, He$^+$, C$^+$, H$_3$O$^+$, HCO$^+$, and N$_2$H$^+$. 
Reactions with H$^+$ and C$^+$ are assumed to transfer charge, 
while He$^+$ reduces the size of the silicon carbon chain by one carbon atom.  All other 
reactions result in proton transfer. We calculate the collisional rate coefficients of these 
such reactions using the formula \citep{su82}:
   \begin{equation}
   \label{eqn:sc}
    \mathrm{k(T)} = 3.87 \times 10^{-9} (\mu_\mathrm{D}/\mu)^{1/2} (\mathrm{T}/300)^{-1/2} \quad \mathrm{cm^3 s^{-1}} \,, 
    \end{equation}
 where $\mu_D$ is the electric dipole moment of the neutral molecule in Debye and $\mu$ is 
 the reduced mass of the reactants in atomic mass units. 

In RATE22, the only neutral-neutral (NN) reactions included for SiC$_3$ and SiC$_4$ were their reaction 
with O atome which release CO. Following the suggestion by \citet{cer25}, we added reactions 
involving small silicon molecules with the radical species CH, C$_2$H, C$_3$H, and C$_4$H that 
result in chain growth, although we ignore reactions that might grow the neutral chains beyond 
SiC$_8$.   For example, the ratefile includes:

\begin{eqnarray}
  \rm{Si + C_{n}H_{2}} & \rightarrow & \rm{SiC_{n} + H_{2}}      
\end{eqnarray}
for n = 2--8.  
For all but C$_2$H$_2$, we adopt a rate 
coefficient of 10$^{-10}$ cm$^3$ s$^{-1}$.  For the reaction involving acetylene, we use
the rate coefficient measured by \citet{can01} over the temperature range 15-300 K, 
2.6 $\times$ 10$^{-10}$ (T/300)$^{-0.71}$ exp(-29/T) cm$^3$ s$^{-1}$.
Clearly, determination of the rate coefficients 
and products of these reactions is an important topic for further study.

We treat the photodissociation of the carbides, SiC$_n$, n = 3-8, in a simple manner, assuming that the 
Si-C bond breaks preferentially as it is the weakest in the molecule but also allowing smaller 
silicon carbides to form. Thus:

\begin{eqnarray}
  \rm{SiC_{n} + h\nu}  & \rightarrow & \rm{Si + C_{n}} \\      
                       & \rightarrow & \rm{SiC_{n-1} + C} \\      
                       & \rightarrow & \rm{SiC_{n-2} + C_{2}}       
\end{eqnarray}
with branching ratios of 0.70, 0.15 and 0.15, respectively.  This reduces the possibility 
of recycling the original molecule, although it does produce two smaller carbides. We adopt 
a total photodissociation rate of 2.0 $\times$ 10$^{-10}$\textrm{e}$^{-2.3{A_V}}$ \textrm{s}$^{-1}$ 
for n $\ge$ 3.

Due to the potential importance of the SiC$_n^+$ reactions with C$_2$H$_2$ as 
measured by \citet{par92}, we have also included photoionisation of SiC$_n$, a process not 
included in our RATE22 network:

\begin{eqnarray}
 \label{eqn:phion}
  \rm{SiC_{n} + h\nu}  & \rightarrow & \rm{SiC_n^+ + e^-}       
\end{eqnarray}
and assumed a rate coefficient of 1.0 $\times$ 
10$^{-10}$\textrm{e}$^{-2.3{A_V}}$ \textrm{s}$^{-1}$ for n $\ge$ 2.

The protonated ions, SiC$_n$H$^+$, are abundant both due to fast proton transfer reactions 
of SiC$_n$  with H$_3^+$, HCO$^+$,
and N$_2$H$^+$, but more importantly, through the reaction of SiC$_n^+$ with C$_2$H$_2$ 
\citep{par92}.  We take a uniform approach to their dissociative recombination (DR) with electrons
and take branching ratios equal to 1/3 for each product channel:

\begin{eqnarray}
  \rm{SiC_{n}H^+ + e^-}  & \rightarrow & \rm{Si + C_{n}H} \\      
                       & \rightarrow & \rm{SiC_{n-1} + CH} \\      
                       & \rightarrow & \rm{SiC_{n} + H}       
\end{eqnarray}
for n = 5-8, while for n = 9 we adopt:

\begin{eqnarray}
  \rm{SiC_{9}H^+ + e^-}  & \rightarrow & \rm{Si + C_{9}H} \\      
                       & \rightarrow & \rm{SiC_2 + C_7H} \\      
                       & \rightarrow & \rm{SiC_8 + CH}       
\end{eqnarray}
to ensure formation of SiC$_8$ through DR is consistent with that of SiC$_7$. We have 
chosen to terminate the chain at this point and so do not include SiC$_9$ in the 
network.  The adoption of the SiC$_2$ + C$_7$H channel ensures the DR process has the same 
overall rate coefficient and branching ratios as those in eqns. 8--10 while having no effect
on the abundance of the parent species SiC$_2$ and SiC$_n$, n = 3--7.

In total, the network contains over 300 reactions involving the Si-C bond among 40 SiC-bearing 
species, including SiC$_n$, n = 1--8, SiC$_n^+$, n = 1--8, SiC$_n$H$^+$, n = 1--9, SiC$_n$H, 
n = 1--3, and SiC$_n$H$_2^+$, n =  1--3.  The full set of 202 reactions that are either new 
or have rate coefficients amended from RATE22 is given in \autoref{tab:sic_network}. 

\section{Model Results}
\label{results}

\begin{figure}
	\includegraphics[width=\columnwidth]{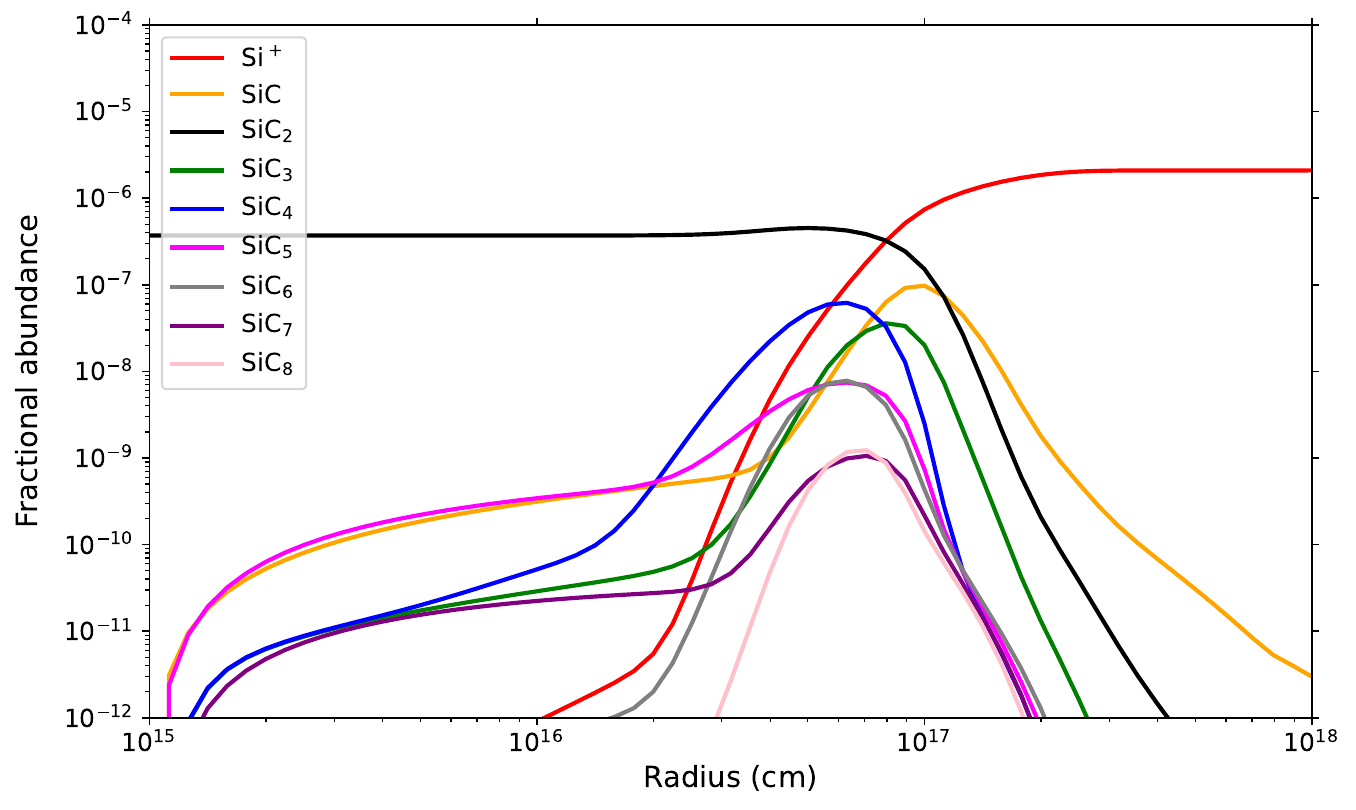}
    \caption{Fractional abundances relative to H$_2$ of the silicon-carbon molecules 
    as a function of radius.}
    \label{fig:fracabund}
\end{figure}

Figure \ref{fig:fracabund} shows the radial distribution of the fractional abundances of the 
silicon-carbon chain species
for our model of IRC+10216 (Model 1).  In common with species formed from
processes initiated by external, interstellar UV photons, fractional abundances reach broad 
peaks over the range (3--7) $\times$ 10$^{16}$ cm, with a general trend that the peak
abundances of SiC$_3$ and SiC$_4$ are about an order of magnitude larger than those of 
SiC$_5$ and SiC$_6$, and about two orders of magnitude larger than those of SiC$_7$ and SiC$_8$.

Observed fractional abundances, however, are derived quantities, often through fitting an
assumed underlying expression for the radial distribution of the species in a radiative transfer 
model that reproduces line profiles and intensities. They may also give 
a false impression of where the specific column density arises; a lower fractional abundance
at smaller radii can have a more significant effect on column density than a higher value at 
larger radii due to the 1/r$^2$ dependence of the H$_2$ abundance.

\begin{figure}
	\includegraphics[width=\columnwidth]{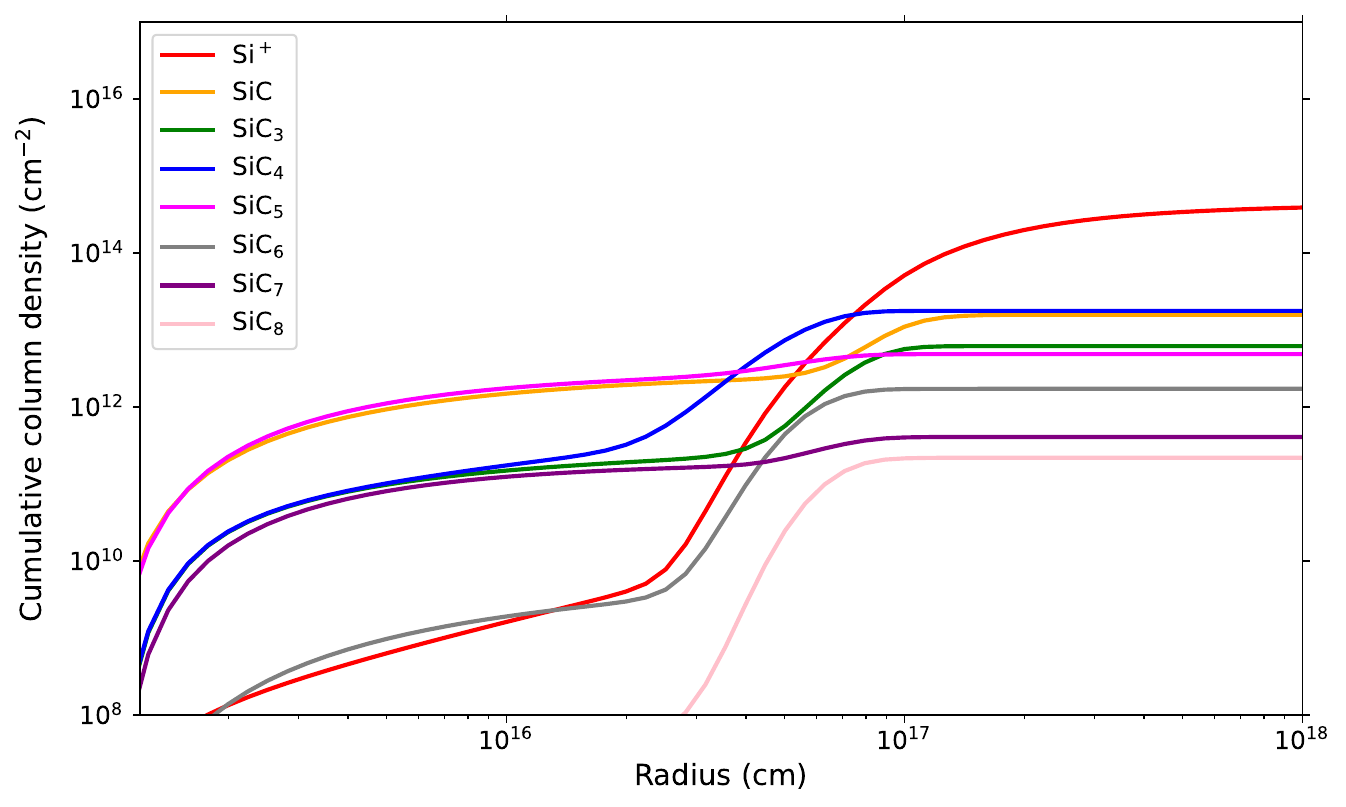}
    \caption{Cumulative column densities of the silicon-carbon molecules 
    as a function of radius.}
    \label{fig:cumcol}
\end{figure}

In Figure \ref{fig:cumcol}, we present the cumulative column densities for these species. 
At their peak radii of maximum fractional abundances, the cumulative column densities
relative to their final values, range from
around 65 percent, for SiC$_3$ and SiC$_4$, to 35--45 percent, for the larger chains. 

Table \ref{tab:coldens} compares the calculated radial column densities (Model 1)
with those observed in IRC+10216.
In this table, we have, for SiC$_3$, adopted the sum of the column densities of its linear 
and rhomboidal forms since we are unable to differentiate these in our model.
Agreement between the model and observations is within a factor of four for 
all species, a rather good fit given the similar values chosen for many of the rate coefficients and 
branching ratios, including those for NN, DR and photoreactions.

\begin{table}
	\centering
	\caption{Comparison between calculated and observed column densities (cm$^{-2}$) of SiC-bearing species in
	IRC+10216.} 
	\label{tab:coldens}
	\begin{tabular}{lcccl} 
		\hline
		Species & Model 1 & Model 2 & Observed & Reference \\
		\hline
		SiC     & 1.59 $\times$ 10$^{13}$ & 1.55 $\times$ 10$^{13}$ & 6.0 $\times$ 10$^{13}$ & \citet{cer89} \\
		SiC$_3$ & 6.24 $\times$ 10$^{12}$ & 3.52 $\times$ 10$^{12}$ & 9.8 $\times$ 10$^{12}$ & \citet{cer25} \\ 
		SiC$_4$ & 1.79 $\times$ 10$^{13}$ & 1.46 $\times$ 10$^{12}$ & 7.3 $\times$ 10$^{12}$ & \citet{par25} \\ 
		SiC$_5$ & 4.91 $\times$ 10$^{12}$ & 2.11 $\times$ 10$^{11}$ & 1.8 $\times$ 10$^{12}$ & \citet{cer25} \\ 
		SiC$_6$ & 1.74 $\times$ 10$^{12}$ & 2.98 $\times$ 10$^{11}$ & 7.9 $\times$ 10$^{11}$ & \citet{par25} \\ 
		SiC$_7$ & 4.21 $\times$ 10$^{11}$ & 7.18 $\times$ 10$^{10}$ & -- & -- \\ 
		SiC$_8$ & 2.23 $\times$ 10$^{11}$ & 8.41 $\times$ 10$^{10}$ & -- & -- \\ 
		\hline
	\end{tabular}
\end{table}

\section{Discussion}
\label{discussion}

\begin{figure}
	\includegraphics[width=\columnwidth]{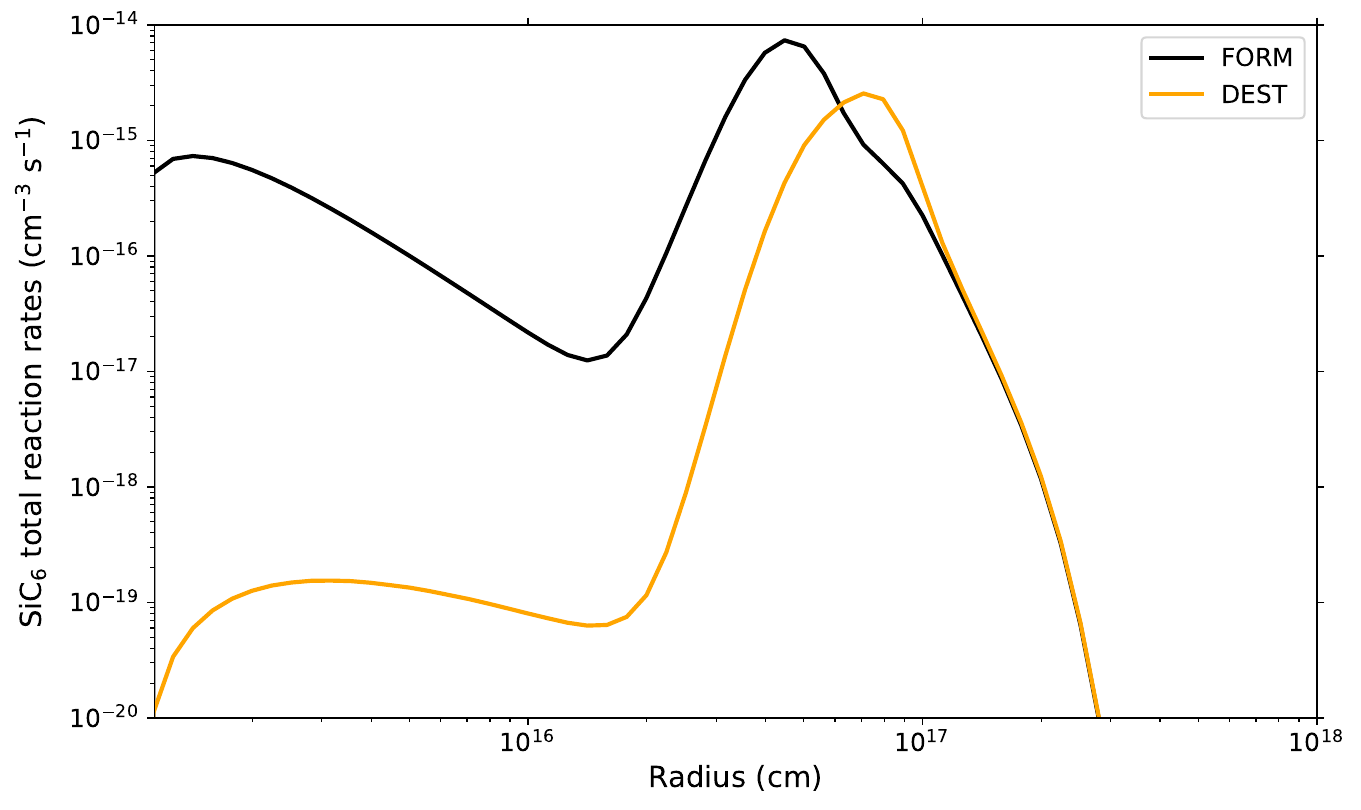}
    \caption{The total formation and destruction rate (cm$^{-3}$ s$^{-1}$) of SiC$_6$ 
    as a function of radius. }
    \label{fig:total}
\end{figure}

To illustrate the chemistry of the new silicon-carbide chain species included in the model, we 
consider the important formation and destruction reactions of SiC$_6$. 
Figure \ref{fig:total} show the overall formation and destruction rates of SiC$_6$ per unit 
volume per second as a function of radial distance.  Inside 10$^{16}$ cm, the destruction 
rate is much smaller than that of formation, primarily due to the lack of ultraviolet photons
in this region.  At 10$^{15}$ cm, A$_V$ = 24.9 mag, n(H$_2$) = 2.33 $\times$ 10$^7$ cm$^{-3}$,
 and T = 286.2 K.  As a result, chemistry is initiated by cosmic ray ionisation of H$_2$ which
leads to proton transfer from H$_3^+$ to abundant neutrals such as CO and C$_2$H$_2$. DR reactions of 
C$_2$H$_3^+$ lead to the reactive radicals C$_2$ and C$_2$H which initiate the build-up of longer
hydrocarbon chain molecules.  Although fractional abundances are small in this region, the 
total formation rates are density dependent, so that their values per cm$^3$ per
second are large.  On the other hand, destruction of molecules such as SiC$_6$ is a slow,
since it primarily depends on interstellar and cosmic ray induced UV photons, both of which have
very low rates in the inner envelope.  Despite the fact that formation is so much larger than loss for 
SiC$_6$, its abundance at r $<$ 10$^{16}$ cm is low because the time-scale for chemical synthesis 
is small - it takes around 200 yr for the flow to reach 10$^{16}$ cm at our adopted expansion
velocity of 14.5 km s$^{-1}$.

Following an initial burst of formation of SiC$_6$, the rate decreases due to the fall in total 
density as the flow expands until, at r $>$ 10$^{16}$ cm, the interstellar UV 
flux becomes large enough to drive a photon-dominated chemistry.  These photons also destroy
SiC$_6$ whose loss rate also rises, before the outflow is eventually reduced to atomic 
gas at radii beyond 10$^{17}$ cm.

\begin{figure}
\subfloat{
	\includegraphics[width=\columnwidth]{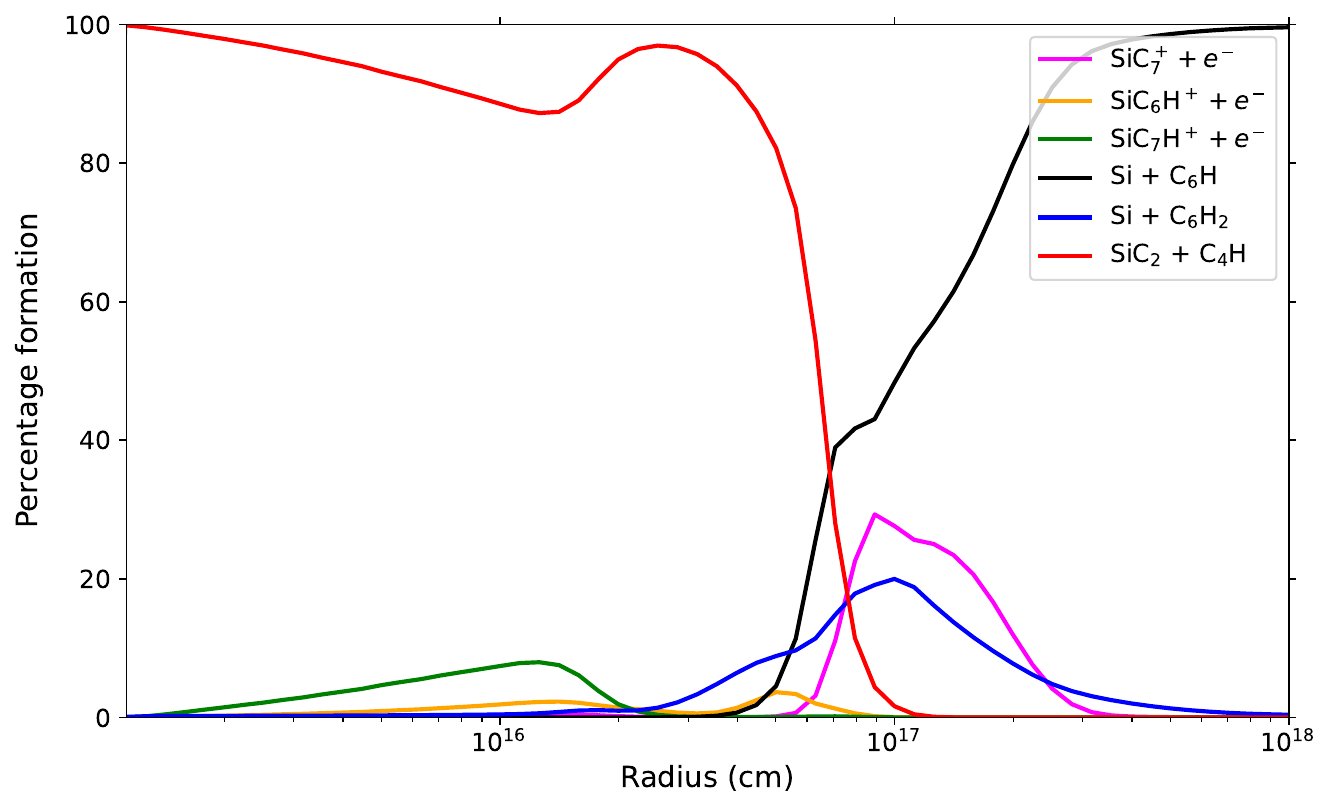}
	}
	\hfill
\subfloat{
	\includegraphics[width=\columnwidth]{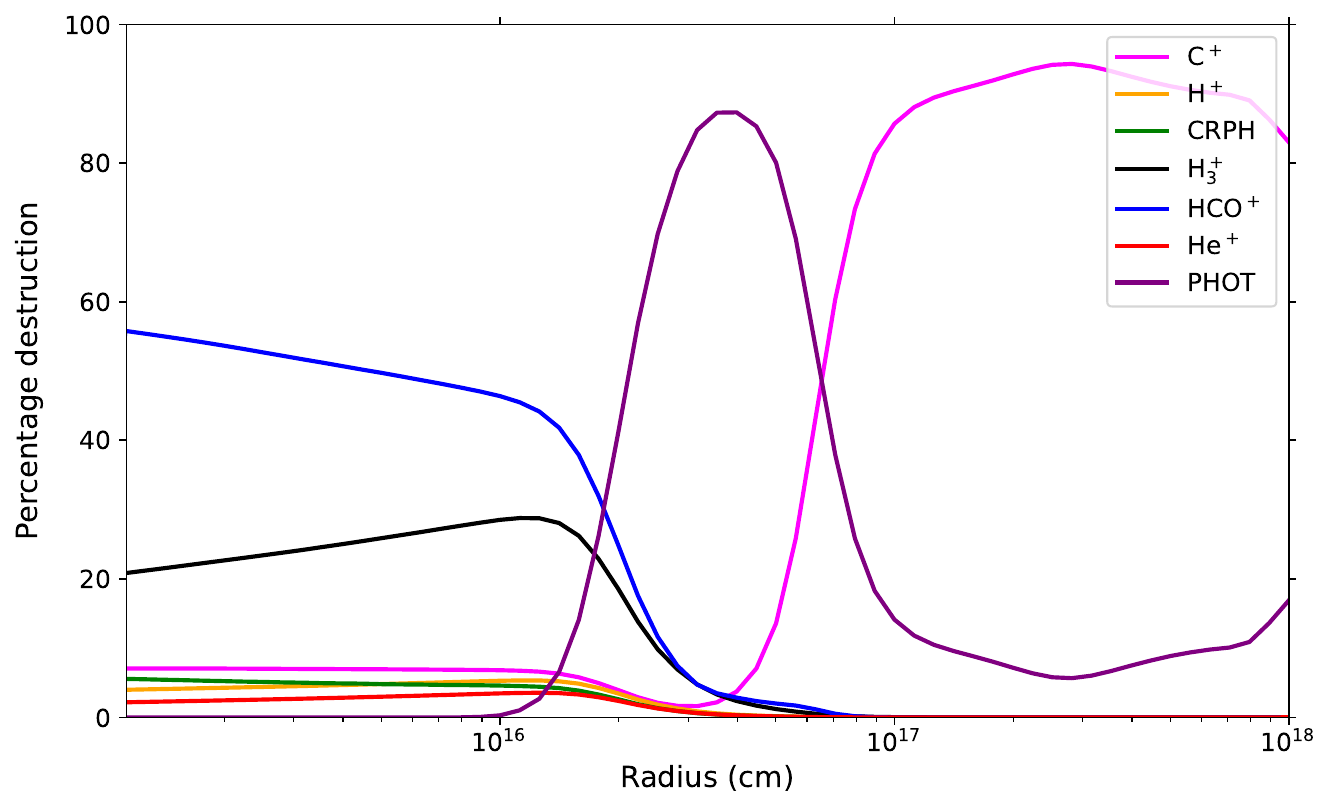}
	}
    \caption{Percentage of the total formation  (upper panel) and destruction
    (lower panel) rates of SiC$_6$ 
    as a function of radius. See text for an explanation of the reaction labels.}
    \label{fig:panels}
\end{figure}

Figure \ref{fig:panels} (upper panel) 
shows the percentage contributions of the six reactions that contribute more than 1 percent 
to the formation of SiC$_6$ as a function of radius. 
The SiC$_6$ fractional abundance reaches a peak value of 7.8 $\times$ 10$^{-9}$ at a 
radius of 6.3 $\times$ 10$^{16}$ cm, at which point the main formation routes are via 
SiC$_2$ + C$_4$H (54 percent), Si + C$_6$H (26 percent), and Si + C$_6$H$_2$ (11 percent).  
At this radius, the Si and SiC$_2$ abundances are almost equal, while that of C$_4$H is 
larger than those of C$_6$H and C$_6$H$_2$.

In Fig. \ref{fig:panels} (lower panel), we show the percentage contributions of 
individual processes to the loss of SiC$_6$. 
Here we have summed up all cosmic ray induced (CRPH) and interstellar UV
(PHOT) photoreactions, that is three photodissocation channels (eqns. 4--6) and one 
photoionisation channel (eqn. \ref{eqn:phion}) for each.  
Proton transfer from HCO$^+$ and H$_3^+$ dominates the low loss rate in the inner 
CSE, interstellar photons dominate in the region of peak fractional abundance and finally 
C$^+$ ions dominate destruction of SiC$_6$ beyond 10$^{17}$ cm.

Finally, we turn to the role that SiC$_2$, which has an essentially unchanged fractional
abundance out to $\sim$ 10$^{17}$ cm, plays in chain formation.  As an abundant parent
molecule which we assume to react with C$_n$H (n = 1--6) with a rate coefficient of 10$^{-10}$
cm$^3$ s$^{-1}$, one might suppose that it is significant. To check this point, we also ran a model 
in which all these rate coefficients were set to zero (Model 2).  The resulting column 
densities of the silicon carbides are shown in Table \ref{tab:coldens}.

\begin{figure}
	\includegraphics[width=\columnwidth]{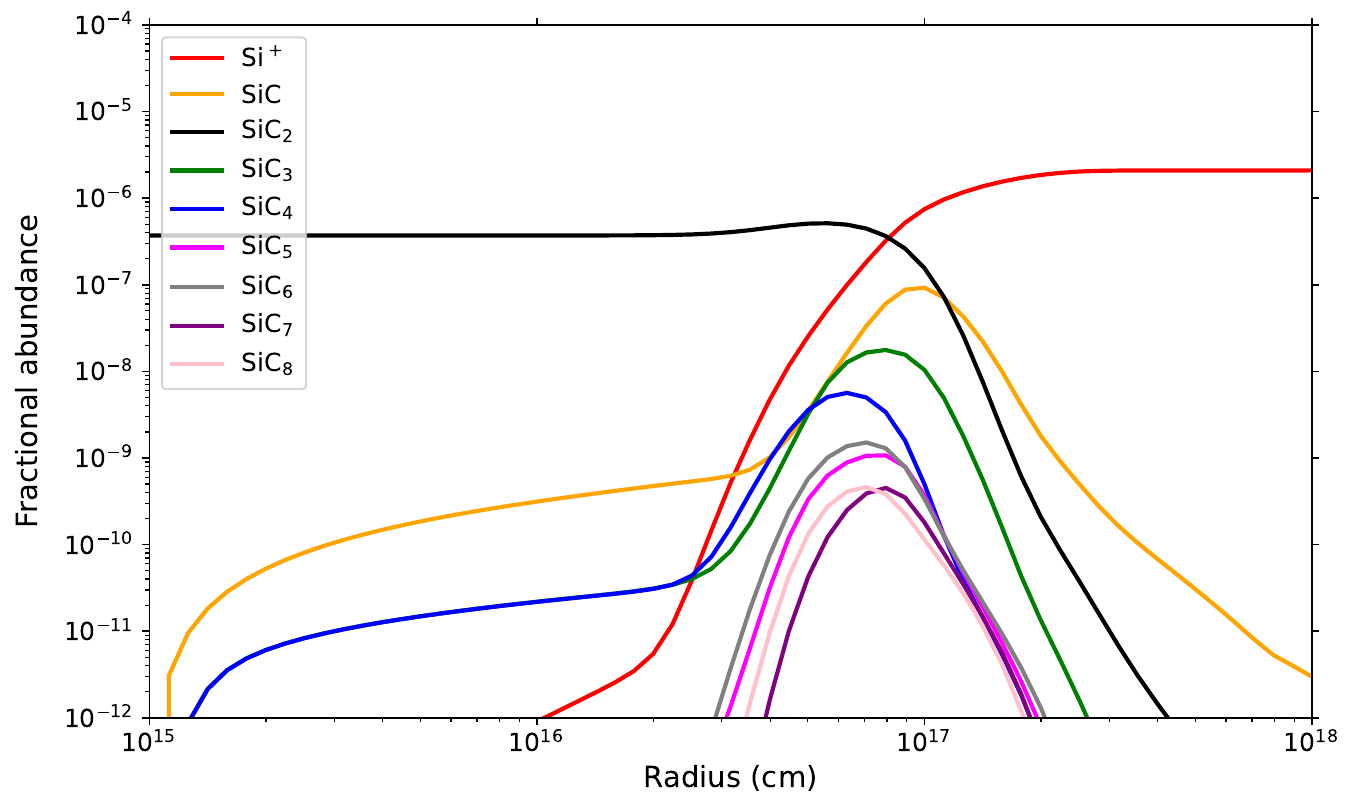}
    \caption{Fractional abundances relative to H$_2$ of the silicon-carbon molecules 
    as a function of radius for the model in which the SiC$_2$ reactions with C$_n$H,
    n = 1--6, are switched off (Model 2).}
    \label{fig:fraczero}
\end{figure}

Fig. \ref{fig:fraczero} shows the resulting radial distribution of the fractional abundances
and indicates that those of SiC$_n$, n $>$ 4, are severely reduced to values less than
10$^{-12}$ inside 4 $\times$ 10$^{16}$ cm.  These low fractional abundances translate into 
lower column densities, up to a factor of about six for SiC$_6$ and SiC$_7$ 
(Table \ref{tab:coldens}). 
Despite this, the reduced column densities all lie within a factor 
of five of those observed.  The one exception is SiC$_5$ which is 
almost an order of magnitude below its observationally determined value \citep{cer25}.

To make an estimate of the detectability of SiC$_7$ and SiC$_8$, we assume that the 
calculated column density ratios of SiC$_7$ to SiC$_5$ and SiC$_8$ to SiC$_6$ replicate 
their ratios in IRC+10216.
We followed the approach
taken by \citet{mil26} to find that the integrated line intensity ratio of SiC$_7$ to
SiC$_5$ is between 0.06 and 0.24 for Models 1 and 2, respectively, and, for SiC$_8$ to SiC$_6$,
between 0.1 and 0.2 for the two models.  Typical integrated line intensities observed in
IRC+10216 are around 17 mK km s$^{-1}$ for SiC$_5$ \citep{cer25} and 19 mK km s$^{-1}$ for
SiC$_6$ \citep{par25}, indicating that the integrated line intensities for the larger carbides 
will be in the level of 1--4 mK km s$^{-1}$.

\section{Conclusions}
\label{conclusions}

We have developed a chemical kinetic model to describe the synthesis of the linear silicon carbide 
molecules in the outer CSE of the carbon-rich AGB star IRC+10216 in an effort to 
reproduce the column densities observed for chains up to SiC$_6$ and to predict those expected for 
SiC$_7$ and SiC$_8$.  To this end, we have included the results of a previously overlooked
experimental study by \citet{par92} on the reactions of SiC$_n^+$ and acetylene. We have also
followed suggestions by \citet{cer25} that growth of these chain molecules involves the
reactions of neutral hydrocarbon radicals such as CH, C$_2$H and C$_4$H with small silicon carbide molecules.

In the absence of any experimental or theoretical evidence on the latter processes, we have adopted a rather simplistic 
approach to rate coefficients and branching ratios for NN, DR and photoreactions. Somewhat
surprisingly, we found that the observed column densities of SiC, and SiC$_3$ to SiC$_6$ are
reproduced to within a factor of about 4.  We predict column densities for SiC$_7$ and SiC$_8$
that are factors of 2-4 less that that observed for SiC$_6$. Given the presence of SiC$_2$
as a parent molecule we have also looked at its influence in forming the silicon carbide chains.  
When the SiC$_2$ reactions with C$_n$H are switched off, fractional abundances 
and column densities for the largest species are reduced but all remain within a factor of 10
of those observed.  As \citet{mcc00} have 
measured the microwave spectra and determined the spectroscopic constants of SiC$_7$ and SiC$_8$,
these species are realistic, although challenging, targets for sensitive searches in IRC+10216.

In our adopted chemical network, the formation of the silicon carbide 
chains is dominated by reactions of Si atoms and SiC$_2$ with neutral hydrocarbons
(Figure \ref{fig:panels}, upper panel). We encourage theoretical and experimental studies of these systems 
to better inform our understanding of silicon-carbon chemistry in IRC+10216.
Finally, we note that it is our intention to include the detailed reaction networks used 
here and in our study of the calcium-bearing cyanopolyynes \citep{mil26}, in the next release 
of the UMIST Database for Astrochemistry, due for the first half of 2027.

\section*{Acknowledgements}

I thank the anonymous reviewer for a timely and helpful report which 
has improved the manuscript. 
TJM's research at Queen's University Belfast is supported by grant ST/T000198/1 from the STFC.

\section*{Data Availability}

 An electronic version of \autoref{tab:sic_network} can 
 be obtained upon request to the author.



\bibliographystyle{mnras}
\bibliography{silicon} 




\appendix
\section{Detailed Silicon Carbide Reaction Network}

In the Appendix, we present the reaction network adopted for the silicon carbide chain chemistry
in Table~\ref{tab:sic_network}.
In this, we have included some reactions whose rate coefficients have been updated since
RATE22, and new reactions as described in Sec. \ref{sec:network}. Numerical values
of the rate coefficients are calculated from the $\alpha$, $\beta$ and $\gamma$  factors
using formulae that depend on the reaction type, as discussed in \citet{mil24}.

\begin{table*}
\caption{Silicon carbide chemical network. 'R' and 'P' refer to reactants and products, 
respectively, `SC' to rate coefficients calculated 
using eqn. \ref{eqn:sc} and `BR' stands for branching ratio. References are: 
(1) \citet{her89}; (2) \citet{par92}; (3) \citet{men24}.}
\label{tab:sic_network}
\begin{tabular}{llllllllllll}\hline
No. & R1 & R2 & P1 & P2 & P3 & P4 & $\alpha$ & $\beta$ & $\gamma$ & Notes & Ref. \\
\hline
1 & C+ & SiC & SiC+ & C & & & 1.09e-09 & -0.50 & 0.0 & SC & 1 \\
2 & C+ & SiC2 & SiC2+ & C & & & 2.96e-09 & -0.50 & 0.0 & SC & 1 \\
3 & C+ & SiC2H & SiC2H+ & C & & & 8.65e-10 & -0.50 & 0.0 & SC & 1 \\
4 & C+ & SiC3 & SiC3+ & C & & & 2.92e-09 & -0.50 & 0.0 & SC & 1 \\
5 & C+ & SiC4 & SiC4+ & C & & & 7.72e-09 & -0.50 & 0.0 & SC & 1 \\
6 & C+ & SiC5 & SiC5+ & C & & & 7.86e-09 & -0.50 & 0.0 & SC & \\
7 & C+ & SiC6 & SiC6+ & C & & & 9.77e-09 & -0.50 & 0.0 & SC & \\
8 & C+ & SiC7 & SiC7+ & C & & & 8.58e-09 & -0.50 & 0.0 & SC & \\
9 & C+ & SiC8 & SiC8+ & C & & & 1.18e-08 & -0.50 & 0.0 & SC & \\
10 & H+ & SiC & SiC+ & H & & & 6.66e-09 & -0.50 & 0.0 & SC & 1 \\
11 & H+ & SiC2 & SiC2+ & H & & & 9.34e-09 & -0.50 & 0.0 & SC & 1 \\                                                                                                                                                                    
12 & H+ & SiC2H2 & SiC2H2+ & H & & & 1.80e-09 & -0.50 & 0.0 & SC & 1 \\                                                                                                                                                                  
13 & H+ & SiC2H & SiC2H+ & H & & & 2.74e-09 & -0.50 & 0.0 & SC & 1 \\                                                                                                                                                                   
14 & H+ & SiC3 & SiC3+ & H & & & 1.87e-08 & -0.50 & 0.0 & SC & 1 \\                                                                                                                                                                      
15 & H+ & SiC3H & SiC3H+ & H & & & 1.50e-09 & -0.50 & 0.0 & SC & 1 \\                                                                                                                                                                    
16 & H+ & SiC4 & SiC4+ & H & & & 2.50e-08 & -0.50 & 0.0 & SC & 1 \\                                                                                                                                                                      
17 & H+ & SiC5 & SiC5+ & H & & & 2.57e-08 & -0.50 & 0.0 & SC & \\                                                                                                                                                                                                                                   
18 & H+ & SiC6 & SiC6+ & H & & & 3.21e-08 & -0.50 & 0.0 & SC & \\                                                                                                                                                                                                                                   
19 & H+ & SiC7 & SiC7+ & H & & & 2.84e-08 & -0.50 & 0.0 & SC & \\                                                                                                                                                                                                                                   
20 & H+ & SiC8 & SiC8+ & H & & & 3.92e-08 & -0.50 & 0.0 & SC \\                                                                                                                                                                                                                                   
21 & SiC2 & CRPHOT & SiC2+ & e- & & & 1.30e-17 & 0.00 & 250 &  & \\                                                                                                                                                                                                                                       
22 & SiC3 & CRPHOT & Si & C3 & & & 1.30e-17 & 0.00 & 250 & BR = 1/3 & \\                                                                                                                                                                                                                                   
23 & SiC3 & CRPHOT & SiC & C2 & & & 1.30e-17 & 0.00 & 250 & BR = 1/3 & \\                                                                                                                                                                                                                                  
24 & SiC3 & CRPHOT & SiC2 & C & & & 1.30e-17 & 0.00 & 250 & BR = 1/3 & \\                                                                                                                                                                                                                                  
25 & SiC3 & CRPHOT & SiC3+ & e- & & & 1.30e-17 & 0.00 & 250 &  & \\                                                                                                                                                                                                                                       
26 & SiC4 & CRPHOT & Si & C4 & & & 1.30e-17 & 0.00 & 250 & BR = 1/3 &  \\                                                                                                                                                                                                                                   
27 & SiC4 & CRPHOT & SiC3 & C & & & 1.30e-17 & 0.00 & 250 & BR = 1/3 &  \\                                                                                                                                                                                                                                  
28 & SiC4 & CRPHOT & SiC2 & C2 & & & 1.30e-17 & 0.00 & 250 & BR = 1/3\\                                                                                                                                                                                                                                 
29 & SiC4 & CRPHOT & SiC4+ & e- & & & 1.30e-17 & 0.00 & 250 &  & \\                                                                                                                                                                                                                                       
30 & SiC5 & CRPHOT & Si & C5 & & & 1.30e-17 & 0.00 & 250 & BR = 1/3 &  \\                                                                                                                                                                                                                                   
31 & SiC5 & CRPHOT & SiC4 & C & & & 1.30e-17 & 0.00 & 250 & BR = 1/3 &  \\                                                                                                                                                                                                                                  
32 & SiC5 & CRPHOT & SiC3 & C2 & & & 1.30e-17 & 0.00 & 250 & BR = 1/3 &  \\                                                                                                                                                                                                                                 
33 & SiC5 & CRPHOT & SiC5+ & e- & & & 1.30e-17 & 0.00 & 250 &  & \\                                                                                                                                                                                                                                        
34 & SiC6 & CRPHOT & Si & C6 & & & 1.30e-17 & 0.00 & 250 & BR = 1/3 &  \\                                                                                                                                                                                                                                  
35 & SiC6 & CRPHOT & SiC5 & C & & & 1.30e-17 & 0.00 & 250 & BR = 1/3 &  \\                                                                                                                                                                                                                                 
36 & SiC6 & CRPHOT & SiC4 & C2 & & & 1.30e-17 & 0.00 & 250 & BR = 1/3 & \\                                                                                                                                                                                                                                 
37 & SiC6 & CRPHOT & SiC6+ & e- & & & 1.30e-17 & 0.00 & 250 & &  \\                                                                                                                                                                                                                                       
38 & SiC7 & CRPHOT & Si & C7 & & & 1.30e-17 & 0.00 & 250 & BR = 1/3 &  \\                                                                                                                                                                                                                                  
39 & SiC7 & CRPHOT & SiC6 & C & & & 1.30e-17 & 0.00 & 250 & BR = 1/3 & \\                                                                                                                                                                                                                                  
40 & SiC7 & CRPHOT & SiC5 & C2 & & & 1.30e-17 & 0.00 & 250 & BR = 1/3 & \\                                                                                                                                                                                                                                 
41 & SiC7 & CRPHOT & SiC7+ & e- & & & 1.30e-17 & 0.00 & 250 &  & \\                                                                                                                                                                                                                                       
42 & SiC8 & CRPHOT & Si & C8 & & & 1.30e-17 & 0.00 & 250 & BR = 1/3 & \\                                                                                                                                                                                                                                   
43 & SiC8 & CRPHOT & SiC7 & C & & & 1.30e-17 & 0.00 & 250 & BR = 1/3 & \\                                                                                                                                                                                                                                  
44 & SiC8 & CRPHOT & SiC6 & C2 & & & 1.30e-17 & 0.00 & 250 & BR = 1/3 & \\                                                                                                                                                                                                                                 
45 & SiC8 & CRPHOT & SiC8+ & e- & & & 1.30e-17 & 0.00 & 250 &  & \\                                                                                                                                                                                                                                       
46 & SiCH4+ & e- & SiCH3 & H & & & 1.50e-07 & -0.50 & 0.0 & & 1 \\                                                                                                                                                
47 & SiC5+ & e- & Si & C5 & & & 1.00e-07 & -0.50 & 0.0 & BR = 1/3 &   \\                                                                                                                                                                                                                                        
48 & SiC5+ & e- & SiC2 & C3 & & & 1.00e-07 & -0.50 & 0.0 & BR = 1/3 &   \\                                                                                                                                                                                                                                      
49 & SiC5+ & e- & SiC4 & C & & & 1.00e-07 & -0.50 & 0.0 & BR = 1/3 &   \\                                                                                                                                                                                                                                       
50 & SiC6+ & e- & Si & C6 & & & 1.00e-07 & -0.50 & 0.0 & BR = 1/3 &   \\                                                                                                                                                                                                                                        
51 & SiC6+ & e- & SiC3 & C3 & & & 1.00e-07 & -0.50 & 0.0 & BR = 1/3 &   \\                                                                                                                                                                                                                                      
52 & SiC6+ & e- & SiC5 & C & & & 1.00e-07 & -0.50 & 0.0 & BR = 1/3 &    \\                                                                                                                                                                                                                                      
53 & SiC7+ & e- & Si & C7 & & & 1.00e-07 & -0.50 & 0.0 & BR = 1/3 &    \\                                                                                                                                                                                                                                       
54 & SiC7+ & e- & SiC4 & C3 & & & 1.00e-07 & -0.50 & 0.0 & BR = 1/3 &   \\                                                                                                                                                                                                                                      
55 & SiC7+ & e- & SiC6 & C & & & 1.00e-07 & -0.50 & 0.0 & BR = 1/3 &   \\                                                                                                                                                                                                                                       
56 & SiC8+ & e- & Si & C8 & & & 1.00e-07 & -0.50 & 0.0 & BR = 1/3 &    \\                                                                                                                                                                                                                                       
57 & SiC8+ & e- & SiC5 & C3 & & & 1.00e-07 & -0.50 & 0.0 & BR = 1/3 &   \\                                                                                                                                                                                                                                      
58 & SiC8+ & e- & SiC7 & C & & & 1.00e-07 & -0.50 & 0.0 & BR = 1/3 &   \\                                                                                                                                                                                                                                       
59 & SiC5H+ & e- & Si & C5H & & & 1.00e-07 & -0.50 & 0.0 & BR = 1/3 & \\                                                                                                                                                                                                                                       
60 & SiC5H+ & e- & SiC5 & H & & & 1.00e-07 & -0.50 & 0.0 & BR = 1/3 & \\                                                                                                                                                                                                                                       
61 & SiC5H+ & e- & SiC4 & CH & & & 1.00e-07 & -0.50 & 0.0 & BR = 1/3 & \\                                                                                                                                                                                                                                      
62 & SiC6H+ & e- & Si & C6H & & & 1.00e-07 & -0.50 & 0.0 & BR = 1/3 & \\                                                                                                                                                                                                                                       
63 & SiC6H+ & e- & SiC6 & H & & & 1.00e-07 & -0.50 & 0.0 & BR = 1/3 &  \\                                                                                                                                                                                                                                      
64 & SiC6H+ & e- & SiC5 & CH & & & 1.00e-07 & -0.50 & 0.0 & BR = 1/3 & \\ 
\hline

\end{tabular}
\end{table*}

\begin{table*}
\contcaption{Silicon carbide chemical network}
\label{tab:continue1}
\begin{tabular}{llllllllllll}\hline
No. & R1 & R2 & P1 & P2 & P3 & P4 & $\alpha$ & $\beta$ & $\gamma$ & Notes & Ref. \\
\hline
65 & SiC7H+ & e- & Si & C7H & & & 1.00e-07 & -0.50 & 0.0 & BR = 1/3 &  \\                                                                                                                                                                                                                                      
66 & SiC7H+ & e- & SiC6 & CH & & & 1.00e-07 & -0.50 & 0.0 & BR = 1/3 &  \\                                                                                                                                                                                                                                     
67 & SiC7H+ & e- & SiC7 & H & & & 1.00e-07 & -0.50 & 0.0 & BR = 1/3 &  \\                                                                                                                                                                                                                                      
68 & SiC8H+ & e- & Si & C8H & & & 1.00e-07 & -0.50 & 0.0 & BR = 1/3 & \\                                                                                                                                                                                                                                       
69 & SiC8H+ & e- & SiC7 & CH & & & 1.00e-07 & -0.50 & 0.0 & BR = 1/3 & \\                                                                                                                                                                                                                                      
70 & SiC8H+ & e- & SiC8 & H & & & 1.00e-07 & -0.50 & 0.0 & BR = 1/3 & \\                                                                                                                                                                                                                                       
71 & SiC9H+ & e- & Si & C9H & & & 1.00e-07 & -0.50 & 0.0 & BR = 1/3 & \\                                                                                                                                                                                                                                       
72 & SiC9H+ & e- & SiC2 & C7H & & & 1.00e-07 & -0.50 & 0.0 & BR = 1/3 & \\                                                                                                                                                                                                                                     
73 & SiC9H+ & e- & SiC8 & CH  & & & 1.00e-07 & -0.50 & 0.0 & BR = 1/3 & \\                                                                                                                                                                                                                                      
74 & SiC4H5+ & e- & Si & C2H2 & C2H2 & H & 1.00e-07 & -0.50 & 0.0 & BR = 1/3 & \\                                                                                                                                                                                                                                
75 & SiC4H5+ & e- & SiC2 & C2H4 & H & & 1.0e-007 & -0.50 & 0.0 & BR = 1/3 & \\                                                                                                                                                                                                                                  
76 & SiC4H5+ & e- & SiC4 & H2 & H2 & H & 1.00e-07 & -0.50 & 0.0 & BR = 1/3 & \\                                                                                                                                                                                                                                  
77 & C+ & SiC2H2 & SiC3H+ & H & & & 1.14e-09 & -0.50 & 0.0 & SC & 1 \\                                                                                                                                                                 
78 & C+ & SiC2H & SiC3+ & H & & & 8.65e-10 & -0.50 & 0.0 & SC & 1 \\                                                                                                                                                                   
79 & C+ & SiC3H & SiC4+ & H & & & 9.40e-10 & -0.50 & 0.0 & SC & 1 \\                                                                                                                                                                   
80 & C+ & SiC & Si+ & C2 & & & 1.09e-09 & -0.50 & 0.0 & SC & \\                                                                                                                                                                      
81 & C2H2 & SiC2+ & SiC4H+ & H & & & 5.16e-10 & 0.00 & 0.0 &  & 2  \\                                                                                                                                                               
82 & C2H2 & SiC2+ & SiC2H+ & C2H & & & 8.40e-11 & 0.00 & 0.0 &  & 2  \\                                                                                                                                                            
83 & C2H2 & SiC3+ & SiC5H+ & H & & & 3.58e-10 & 0.00 & 0.0 &  & 2  \\                                                                                                                                                              
84 & C2H2 & SiC3+ & SiC2H+ & C3H & & & 2.02e-10 & 0.00 & 0.0 & & 2  \\                                                                                                                                                            
85 & C2H2 & SiC4+ & SiC6H+ & H & & & 6.30e-10 & 0.00 & 0.0 &  & 2  \\                                                                                                                                                              
86 & C2H2 & SiC5+ & SiC7H+ & H & & & 4.60e-10 & 0.00 & 0.0 &  & 2   \\                                                                                                                                                             
87 & C2H2 & SiC6+ & SiC8H+ & H & & & 4.70e-10 & 0.00 & 0.0 &  & 2  \\                                                                                                                                                              
88 & C2H2 & SiC7+ & SiC9H+ & H & & & 2.04e-10 & 0.00 & 0.0 &  & 2  \\                                                                                                                                                              
89 & H3+ & SiC & HCSi+ & H2 & & & 3.94e-09 & -0.50 & 0.0 & SC & 1 \\                                                                                                                                                                   
90 & H3+ & SiC2 & SiC2H+ & H2 & & & 5.49e-09 & -0.50 & 0.0 & SC & 1 \\                                                                                                                                                                 
91 & H3+ & SiC2H2 & SiC2H3+ & H2 & & & 2.11e-09 & -0.50 & 0.0 & SC & 1 \\                                                                                                                                                              
92 & H3+ & SiC2H & SiC2H2+ & H2 & & & 3.22e-09 & -0.50 & 0.0 & SC & 1 \\                                                                                                                                                               
93 & H3+ & SiC3 & SiC3H+ & H2 & & & 1.10e-08 & -0.50 & 0.0 & SC & 1 \\                                                                                                                                                                 
94 & H3+ & SiC3H & SiC3H2+ & H2 & & & 1.76e-09 & -0.50 & 0.0 & SC & 1 \\                                                                                                                                                               
95 & H3+ & SiC4 & SiC4H+ & H2 & & & 1.46e-08 & -0.50 & 0.0 & SC & 1 \\                                                                                                                                                                 
96 & H3+ & SiC5 & SiC5H+ & H2 & & & 1.50e-08 & -0.50 & 0.0 & SC &  \\                                                                                                                                                                                                                              
97 & H3+ & SiC6 & SiC6H+ & H2 & & & 1.87e-08 & -0.50 & 0.0 & SC & \\                                                                                                                                                                                                                              
98 & H3+ & SiC7 & SiC7H+ & H2 & & & 1.65e-08 & -0.50 & 0.0 & SC & \\                                                                                                                                                                                                                              
99 & H3+ & SiC8 & SiC8H+ & H2 & & & 2.28e-08 & -0.50 & 0.0 & SC & \\                                                                                                                                                                                                                              
100 & H3O+ & SiC & HCSi+ & H2O & & & 1.83e-09 & -0.50 & 0.0 & SC & 1 \\                                                                                                                                                                 
101 & H3O+ & SiC2 & SiC2H+ & H2O & & & 2.48e-09 & -0.50 & 0.0 & SC & 1 \\                                                                                                                                                               
102 & H3O+ & SiC2H & SiC2H2+ & H2O & & & 1.45e-09 & -0.50 & 0.0 & SC & 1 \\                                                                                                                                                             
103 & H3O+ & SiC2H2 & SiC2H3+ & H2O & & & 9.50e-10 & -0.50 & 0.0 & SC & 1 \\                                                                                                                                                            
104 & H3O+ & SiC3 & SiC3H+ & H2O & & & 4.85e-09 & -0.50 & 0.0 & SC & 1 \\                                                                                                                                                               
105 & H3O+ & SiC3H & SiC3H2+ & H2O & & & 7.80e-10 & -0.50 & 0.0 & SC & 1 \\                                                                                                                                                             
106 & H3O+ & SiC4 & SiC4H+ & H2O & & & 6.37e-09 & -0.50 & 0.0 & SC & 1 \\                                                                                                                                                               
107 & H3O+ & SiC5 & SiC5H+ & H2O & & & 6.46e-09 & -0.50 & 0.0 & SC & \\                                                                                                                                                                                                                            
108 & H3O+ & SiC6 & SiC6H+ & H2O & & & 8.00e-09 & -0.50 & 0.0 & SC & \\                                                                                                                                                                                                                            
109 & H3O+ & SiC7 & SiC7H+ & H2O & & & 7.01e-09 & -0.50 & 0.0 & SC & \\                                                                                                                                                                                                                            
110 & H3O+ & SiC8 & SiC8H+ & H2O & & & 9.63e-09 & -0.50 & 0.0 & SC & \\                                                                                                                                                                                                                            
111 & HCO+ & SiC & HCSi+ & CO & & & 1.60e-09 & -0.50 & 0.0 & SC & 1 \\                                                                                                                                                                  
112 & HCO+ & SiC2 & SiC2H+ & CO & & & 2.14e-09 & -0.50 & 0.0 & SC & 1 \\                                                                                                                                                                
113 & HCO+ & SiC2H2 & SiC2H3+ & CO & & & 8.20e-10 & -0.50 & 0.0 & SC & 1 \\                                                                                                                                                             
114 & HCO+ & SiC2H & SiC2H2+ & CO & & & 1.25e-09 & -0.50 & 0.0 & SC & 1 \\                                                                                                                                                              
115 & HCO+ & SiC3 & SiC3H+ & CO & & & 4.16e-09 & -0.50 & 0.0 & SC & 1 \\                                                                                                                                                                
116 & HCO+ & SiC3H & SiC3H2+ & CO & & & 6.70e-10 & -0.50 & 0.0 & SC & 1 \\                                                                                                                                                              
117 & HCO+ & SiC4 & SiC4H+ & CO & & & 5.42e-09 & -0.50 & 0.0 & SC & 1 \\                                                                                                                                                                   
118 & HCO+ & SiC5 & SiC5H+ & CO & & & 5.47e-09 & -0.50 & 0.0 & SC & \\                                                                                                                                                                                                                             
119 & HCO+ & SiC6 & SiC6H+ & CO & & & 6.74e-09 & -0.50 & 0.0 & SC & \\                                                                                                                                                                                                                             
120 & HCO+ & SiC7 & SiC7H+ & CO & & & 5.89e-09 & -0.50 & 0.0 & SC & \\                                                                                                                                                                                                                             
121 & HCO+ & SiC8 & SiC8H+ & CO & & & 8.06e-09 & -0.50 & 0.0 & SC & \\                                                                                                                                                                                                                             
122 & He+ & SiC2 & Si+ & C2 & He & & 4.80e-09 & -0.50 & 0.0 & SC & 1 \\                                                                                                                                                                  
123 & He+ & SiC2H2 & SiC2H+ & He & H & & 9.20e-10 & -0.50 & 0.0 & SC & 1 \\                                                                                                                                                              
124 & He+ & SiC2H2 & SiH+ & C2H & He & & 9.20e-10 & -0.50 & 0.0 & SC & 1 \\                                                                                                                                                              
125 & He+ & SiC2H & Si+ & C2H & He & & 1.40e-09 & -0.50 & 0.0 & SC & 1 \\                                                                                                                                                                
126 & He+ & SiC2H & SiC2+ & He & H & & 1.40e-09 & -0.50 & 0.0 & SC & 1 \\                                                                                                                                                                
127 & He+ & SiC3 & SiC2+ & C & He &  & 9.57e-09 & -0.50 & 0.0 & SC & 1 \\                                                                                                                                                                 
128 & He+ & SiC3H & C3H & Si+ & He & & 7.70e-10 & -0.50 & 0.0 & SC & 1 \\  
\hline

\end{tabular}
\end{table*}

\begin{table*}
\contcaption{Silicon carbide chemical network}
\label{tab:continue2}
\begin{tabular}{llllllllllll}\hline
No. & R1 & R2 & P1 & P2 & P3 & P4 & $\alpha$ & $\beta$ & $\gamma$ & Notes & Ref. \\
\hline
129 & He+ & SiC3H & SiC3+ & He & H & & 7.70e-09 & -0.50 & 0.0 & SC & 1 \\                                                                                                                                                                
130 & He+ & SiC4 & SiC3+ & C & He & & 1.28e-08 & -0.50 & 0.0 & SC & 1 \\                                                                                                                                                                 
131 & He+ & SiC5 & SiC4+ & C & He & & 1.31e-08 & -0.50 & 0.0 & SC & \\                                                                                                                                                                                                                              
132 & He+ & SiC6 & SiC5+ & C & He & & 1.63e-08 & -0.50 & 0.0 & SC & \\                                                                                                                                                                                                                              
133 & He+ & SiC7 & SiC6+ & C & He & & 1.44e-08 & -0.50 & 0.0 & SC & \\                                                                                                                                                                                                                              
134 & He+ & SiC8 & SiC7+ & C & He & & 1.99e-08 & -0.50 & 0.0 & SC & \\                                                                                                                                                                                                                              
135 & N2H+ & SiC & HCSi+ & N2 & & & 1.60e-09 & 0.00 & 0.0 & SC & \\                                                                                                                                                                                                                                
136 & N2H+ & SiCH2 & SiCH3+ & N2 & & & 2.00e-09 & 0.00 & 0.0 &  & \\                                                                                                                                                                                                                                    
137 & N2H+ & SiC2 & SiC2H+ & N2 & & & 2.14e-09 & -0.50 & 0.0 & SC &  \\                                                                                                                                                                                                                             
138 & N2H+ & SiC2H & SiC2H2+ & N2 & & & 1.25e-09 & -0.50 & 0.0 & SC & \\                                                                                                                                                                                                                           
139 & N2H+ & SiC2H2 & SiC2H3+ & N2 & & & 8.20e-10 & -0.50 & 0.0 & SC & \\                                                                                                                                                                                                                          
140 & N2H+ & SiC3 & SiC3H+ & N2 & & & 4.16e-09 & -0.50 & 0.0 & SC & \\                                                                                                                                                                                                                             
141 & N2H+ & SiC3H & SiC3H2+ & N2 & & & 6.20e-10 & -0.50 & 0.0 & SC & \\                                                                                                                                                                                                                           
142 & N2H+ & SiC4 & SiC4H+ & N2 & & & 5.42e-09 & -0.50 & 0.0 & SC & \\                                                                                                                                                                                                                             
143 & N2H+ & SiC5 & SiC5H+ & N2 & & & 5.47e-09 & -0.50 & 0.0 & SC & \\                                                                                                                                                                                                                             
144 & N2H+ & SiC6 & SiC6H+ & N2 & & & 6.74e-09 & -0.50 & 0.0 & SC & \\                                                                                                                                                                                                                             
145 & N2H+ & SiC7 & SiC7H+ & N2 & & & 5.85e-09 & -0.50 & 0.0 & SC & \\                                                                                                                                                                                                                             
146 & N2H+ & SiC8 & SiC8H+ & N2 & & & 8.06e-09 & -0.50 & 0.0 & SC & \\                                                                                                                                                                                                                             
147 & Si & C3H & SiC3 & H & & & 1.00e-10 & 0.00 & 0.0 &  &  \\                                                                                                                                                                                                                                                 
148 & Si & C4H & SiC4 & H & & & 1.00e-10 & 0.00 & 0.0 &  & \\                                                                                                                                                                                                                                                 
149 & Si & C5H & SiC5 & H & & & 1.00e-10 & 0.00 & 0.0 &  & \\                                                                                                                                                                                                                                                 
150 & Si & C6H & SiC6 & H & & & 1.00e-10 & 0.00 & 0.0 &  & \\                                                                                                                                                                                                                                                 
151 & Si & C7H & SiC7 & H & & & 1.00e-10 & 0.00 & 0.0 &  & \\                                                                                                                                                                                                                                                 
152 & Si & C8H & SiC8 & H & & & 1.00e-10 & 0.00 & 0.0 & &  \\                                                                                                                                                                                                                                                  
153 & Si & H2CCC & SiC3 & H2 & & & 1.00e-10 & 0.00 & 0.0 & &  \\                                                                                                                                                                                                                                               
154 & Si & C4H2 & SiC4 & H2 & & & 1.00e-10 & 0.00 & 0.0 &  & \\                                                                                                                                                                                                                                              
155 & Si & C5H2 & SiC5 & H2 & & & 1.00e-10 & 0.00 & 0.0 &  & \\                                                                                                                                                                                                                                               
156 & Si & C6H2 & SiC6 & H2 & & & 1.00e-10 & 0.00 & 0.0 &  & \\                                                                                                                                                                                                                                               
157 & Si & C7H2 & SiC7 & H2 & & & 1.00e-10 & 0.00 & 0.0 &  & \\                                                                                                                                                                                                                                               
158 & Si & C8H2 & SiC8 & H2 & & & 1.00e-10 & 0.00 & 0.0 &  & \\                                                                                                                                                                                                                                               
159 & Si & CH & SiC & H & & & 1.00e-10 & 0.00 & 0.0 &  & \\                                                                                                                                                                                                                                               
160 & SiC2 & CH & SiC3 & H & & & 1.00e-10 & 0.00 & 0.0 &  & \\                                                                                                                                                                                                                                                 
161 & SiC3 & CH & SiC4 & H & & & 1.00e-10 & 0.00 & 0.0 &  & \\                                                                                                                                                                                                                                                
162 & SiC4 & CH & SiC5 & H & & & 1.00e-10 & 0.00 & 0.0 &  & \\                                                                                                                                                                                                                                                
163 & SiC5 & CH & SiC6 & H & & & 1.00e-10 & 0.00 & 0.0 &  & \\                                                                                                                                                                                                                                                
164 & SiC6 & CH & SiC7 & H & & & 1.00e-10 & 0.00 & 0.0 &  & \\                                                                                                                                                                                                                                               
165 & SiC7 & CH & SiC8 & H & & & 1.00e-10 & 0.00 & 0.0 &  & \\                                                                                                                                                                                                                                                
166 & Si & C2H & SiC2 & H & & & 1.00e-10 & 0.00 & 0.0 &  & \\                                                                                                                                                                                                                                                  
167 & SiC & C2H & SiC3 & H & & & 1.00e-10 & 0.00 & 0.0 &  & \\                                                                                                                                                                                                                                                 
168 & SiC2 & C2H & SiC4 & H & & & 1.00e-10 & 0.00 & 0.0 &  & \\                                                                                                                                                                                                                                                
169 & SiC2 & C3H & SiC5 & H & & & 1.00e-10 & 0.00 & 0.0 &  & \\                                                                                                                                                                                                                                               
170 & SiC2 & C4H & SiC6 & H & & & 1.00e-10 & 0.00 & 0.0 &  & \\                                                                                                                                                                                                                                               
171 & SiC2 & C5H & SiC7 & H & & & 1.00e-10 & 0.00 & 0.0 &  & \\                                                                                                                                                                                                                                               
172 & SiC2 & C6H & SiC8 & H & & & 1.00e-10 & 0.00 & 0.0 &  & \\                                                                                                                                                                                                                                                
173 & C & SiS & S & SiC & & & 1.20e-09 & 0.02 & 23261 &  & 3 \\                                                                                                                                                                                  
174 & S & SiC & C & SiS & & & 2.81e-10 & -0.171 & -0.38 &  & 3  \\                                                                                                                                                                                 
175 & S & SiC & Si & CS & & & 3.20e-10 & -0.152 & 2.07 &  &  3 \\                                                                                                                                                                                 
176 & Si & CS & S & SiC & & & 3.84e-09 & 0.143 & 35604 &  &  3 \\                                                                                                                                                                                 
177 & SiC3 & PHOTON & Si & C3 & & & 1.40e-10 & 0.00 & 2.3 & BR = 0.70 & \\                                                                                                                                                                                                                                     
178 & SiC3 & PHOTON & C2 & SiC & & & 3.00e-11 & 0.00 & 2.3 & BR = 0.15 & \\                                                                                                                                                                                                                                    
179 & SiC3 & PHOTON & SiC2 & C & & & 3.00e-11 & 0.00 & 2.3 & BR = 0.15 & \\                                                                                                                                                                                                                                   
180 & SiC3 & PHOTON & SiC3+ & e- & & & 1.00e-10 & 0.00 & 2.3 &  & \\                                                                                                                                                                                                                                   
181 & SiC3H & PHOTON & Si & C3H & & & 1.40e-10 & 0.00 & 2.3 & BR = 0.70 & \\                                                                                                                                                                                                                                  
182 & SiC3H & PHOTON & SiC3 & H & & & 6.00e-11 & 0.00 & 2.3 & BR = O.30 & \\                                                                                                                                                                                                                                  
183 & SiC4 & PHOTON & Si & C4 & & & 1.40e-10 & 0.00 & 2.3 & BR = 0.70 & \\                                                                                                                                                                                                                                  
184 & SiC4 & PHOTON & SiC3 & C & & & 3.00e-11 & 0.00 & 2.3 & BR = 0.15 & \\                                                                                                                                                                                                                                    
185 & SiC4 & PHOTON & SiC2 & C2 & & & 3.00e-11 & 0.00 & 2.3 & BR = 0.15 & \\                                                                                                                                                                                                                                   
186 & SiC4 & PHOTON & SiC4+ & e- & & & 1.00e-10 & 0.00 & 2.3 &  & \\                                                                                                                                                                                                                                  
187 & SiC5 & PHOTON & Si & C5 & & & 1.40e-10 & 0.00 & 2.3 & BR = 0.70 & \\                                                                                                                                                                                                                                     
188 & SiC5 & PHOTON & SiC4 & C & & & 3.00e-11 & 0.00 & 2.3 & BR = 0.15 & \\                                                                                                                                                                                                                                    
189 & SiC5 & PHOTON & SiC3 & C2 & & & 3.00e-11 & 0.00 & 2.3 & BR = 0.15 & \\                                                                                                                                                                                                                                   
190 & SiC5 & PHOTON & SiC5+ & e- & & & 1.00e-10 & 0.00 & 2.3 &  & \\                                                                                                                                                                                                                                  
191 & SiC6 & PHOTON & Si & C6 & & & 1.40e-10 & 0.00 & 2.3 & BR = 0.70 & \\                                                                                                                                                                                                                                     
192 & SiC6 & PHOTON & SiC5 & C & & & 3.00e-11 & 0.00 & 2.3 & BR = 0.15 & \\                                                                                                                                                                                                                                    
\hline

\end{tabular}
\end{table*}

\begin{table*}
\contcaption{Silicon carbide chemical network}
\label{tab:continue3}
\begin{tabular}{llllllllllll}\hline
No. & R1 & R2 & P1 & P2 & P3 & P4 & $\alpha$ & $\beta$ & $\gamma$ & Notes & Ref. \\
\hline
193 & SiC6 & PHOTON & SiC4 & C2 & & & 3.00e-11 & 0.00 & 2.3 & BR = 0.15 & \\                                                                                                                                                                                                                                   
194 & SiC6 & PHOTON & SiC6+ & e- & & & 1.00e-10 & 0.00 & 2.3 &  & \\                                                                                                                                                                                                                                  
195 & SiC7 & PHOTON & Si & C7 & & & 1.40e-10 & 0.00 & 2.3 & BR = 0.70 & \\                                                                                                                                                                                                                                     
196 & SiC7 & PHOTON & SiC6 & C & & & 3.00e-11 & 0.00 & 2.3 & BR = 0.15 & \\                                                                                                                                                                                                                                    
197 & SiC7 & PHOTON & SiC5 & C2 & & & 3.00e-11 & 0.00 & 2.3 & BR = 0.15 & \\                                                                                                                                                                                                                                   
198 & SiC7 & PHOTON & SiC7+ & e- & & & 1.00e-10 & 0.00 & 2.3 &  & \\                                                                                                                                                                                                                                  
199 & SiC8 & PHOTON & Si & C8 & & & 1.40e-10 & 0.00 & 2.3 & BR = 0.70 & \\                                                                                                                                                                                                                                    
200 & SiC8 & PHOTON & SiC7 & C & & & 3.00e-11 & 0.00 & 2.3 & BR = 0.15 & \\                                                                                                                                                                                                                                    
201 & SiC8 & PHOTON & SiC6 & C2 & & & 3.00e-11 & 0.00 & 2.3 & BR = 0.15 & \\                                                                                                                                                                                                                                   
202 & SiC8 & PHOTON & SiC8+ & e- & & & 1.00e-10 & 0.00 & 2.3 & &  \\                                                                                                                                                                                                                                  
\hline

\end{tabular}
\end{table*}


\bsp	
\label{lastpage}
\end{document}